\documentclass[prl,twocolumn,showpacs,superscriptaddress,preprintnumbers,amsmath,amssymb,floatfix,10pt]{revtex4-1}

\newcommand \beq {\begin{equation}}
\newcommand \eeq {\end{equation}}
\newcommand \ben {\begin{eqnarray}}
\newcommand \een {\end{eqnarray}}

\usepackage{wrapfig}
\usepackage{times}
\usepackage{epsfig}
\usepackage{graphicx}
\usepackage{color}
\usepackage{dcolumn}
\usepackage{bm}
\usepackage{amsmath}
\usepackage{setspace}

\begin{document}

\title{Free energy functionals for efficient phase field crystal
  modeling of structural phase transformations}

\author{Michael Greenwood}
\affiliation{Department of Physics and Astronomy, University of
British Columbia, 6224 Agricultural Road Vancouver, BC, V6T 1Z1,
Canada} \affiliation{Department of Materials Science and
Engineering, McMaster University, 1280 Main Street West, Hamilton,
Ontario, L8S 4L7, Canada}

\author{Nikolas Provatas}
\affiliation{Department of Materials Science and Engineering,
McMaster University, 1280 Main Street West, Hamilton, Ontario, L8S
4L7, Canada}

\author{J\"org Rottler}
\affiliation{Department of Physics and Astronomy, University of
British Columbia, 6224 Agricultural Road Vancouver, BC, V6T 1Z1,
Canada}

\begin{abstract}
The phase field crystal (PFC) method has emerged as a promising
technique for modeling materials with atomistic resolution on
mesoscopic time scales. The approach is numerically much more
efficient than classical density functional theory (CDFT), but its
single mode free energy functional only leads to lattices with
triangular (2D) or BCC (3D) symmetries. By returning to a closer
approximation of the CDFT free energy functional, we develop a
systematic construction of two-particle direct correlation functions
that allow the study of a broad class of crystalline structures.
This construction examines planar spacings, lattice symmetries,
planar atomic densities and the atomic vibrational amplitude in the
unit cell of the lattice and also provides control parameters for
temperature and anisotropic surface energies. The power of this new
approach is demonstrated by two examples of structural phase
transformations.
\end{abstract}


\maketitle

Solid-state transformations form the basis of many problems in
materials science and physical metallurgy \cite{Martin1997}.  They
involve complex structural changes between parent and daughter
phases and couple atomic-scale elastic and plastic effects with
diffusional processes. These phenomena are presently impossible to
compute at experimentally relevant time scales using molecular
dynamics simulations.  On the other hand, meso-scale continuum
models wash out most of the relevant atomic scale physics that leads
to elasticity, plasticity, defect interactions and grain boundary
nucleation and migration. Phase field studies of precipitate and
ledge growth \cite{Yeon2005,Loginova2004,Greenwood2009} must thus
re-introduce these effects phenomenologically.  There is presently
no continuum method to efficiently simulate solid-state
transformations on diffusional time scales that self-consistently
computes elastic and plastic effects at the atomic scale.

Classical density function theory (CDFT) provides a formalism that
can accurately describe the emergence of crystalline order from a
liquid or solid phase through a coarse-grained density field
\cite{Ramakrishnan1979}.  Unfortunately, this approach requires very
high spatial resolution and is highly inefficient for dynamical
calculations \cite{Jaatinen2009}. A recent model, coined the phase
field crystal (PFC) model, has been gaining widespread recognition
as a hybrid method between CDFT and traditional phase field methods.
PFC models capture most of the essential physics of CDFT without
having to resolve atomically sharp atomic density peaks
\cite{Elder2002,Elder2004,Elder2007,Stefanovic2006,Mellenthin2008}.
In spite of their successes, however, only PFC free energy
functionals minimized by triangular (2D) or BCC (3D) lattices have
been seriously studied.  While there have been attempts to extend
the PFC model to describe other crystal symmetries such as square
\cite{KroegerThesis2005,Tupper2007} and FCC
\cite{WuThesis2006,Tegze2009,Jaatinen2009} lattices, these have been
somewhat ad-hod and not self-consistently connected to material
properties. PFC modeling is presently lacking a generalized free
energy formulation that allows the study of important phase
transformations between {\it different} crystalline states.

This Letter proposes a new CDFT/PFC model based on two-point direct
correlation function that is systematically constructed to be
minimized by arbitrary crystalline states.  The utility and
efficiency of our density functional model is demonstrated by
dynamically simulating the growth of solid phases into a liquid and
the nucleation of precipitate phases within a parent phase of
different crystallographic symmetry. We begin with the CDFT free
energy functional, \ben \Delta F[n({\vec r})] =\Delta F_{id}[n({\vec
    r})]+\Delta F_{ex}[n({\vec r})],\label{ModelEnergy}\een
where $\Delta F$ is the free energy difference with respect to a
reference state at density $\rho_0$ \cite{Ramakrishnan1979}, and has
contributions from a non-interacting term $\Delta F_{id}$ and an
excess energy $\Delta F_{ex}$. The latter is responsible for the
formation of structured phases.

Eq.~(\ref{ModelEnergy}) is expanded around the reference density
using the dimensionless number density $n(\vec{r}) =
\rho(\vec{r})/\rho_o-1$, where $\rho(\vec{r})$ is the local density.
The non-interacting contribution is approximated by expanding the
Helmholtz free energy of an ideal gas, \ben\Delta F_{id} = \rho_o
k_B T\int d\vec{r} (1+ n(\vec{r})) ln(1+n(\vec{r}))-n(\vec{r})
\nonumber
\\ \approx \rho_o k_B T\int d\vec{r} \left[\frac{n(\vec{r})^2}{2}
-\frac{n(\vec{r})^3}{6}+ \frac{n(\vec{r})^4}{12}\right]. \een
 \noindent
The excess energy is expanded as in CDFT \cite{Ramakrishnan1979} and
earlier PFC models \cite{Teeffelen2009,Elder2007} up to second order
correlations, \ben \Delta F_{ex} = -\frac{\rho_o k_B T}{2} \int d
\vec{r} \, n(\vec{r}) \int d\vec{r}'\left(
C_2(|r-r'|)n(\vec{r}')\right), \label{Fex} \een \noindent where
$C_2(|\vec{r}-\vec{r'}|)$ is the two particle direct correlation
function.  The dimensionless number density field $ n(\vec{r})$ is
evolved in time using conserved dissipative dynamics, \ben
\frac{\partial n(\vec{r})}{\partial t} =M \nabla^2 \frac{\delta
\Delta F}{\delta n(\vec{r})} + \eta\label{Dyn:Dynamics} \een
\noindent where $M$ is a kinetic mobility parameter and $\eta$ is
conserved random noise.  For increased numerical efficiency,
Eq.~(\ref{Fex}) is integrated in reciprocal space using a
semi-implicit technique \cite{Mellenthin2008,Elder2004} .


\begin{figure}[t]
    \centerline{\includegraphics*[width=3.4in]{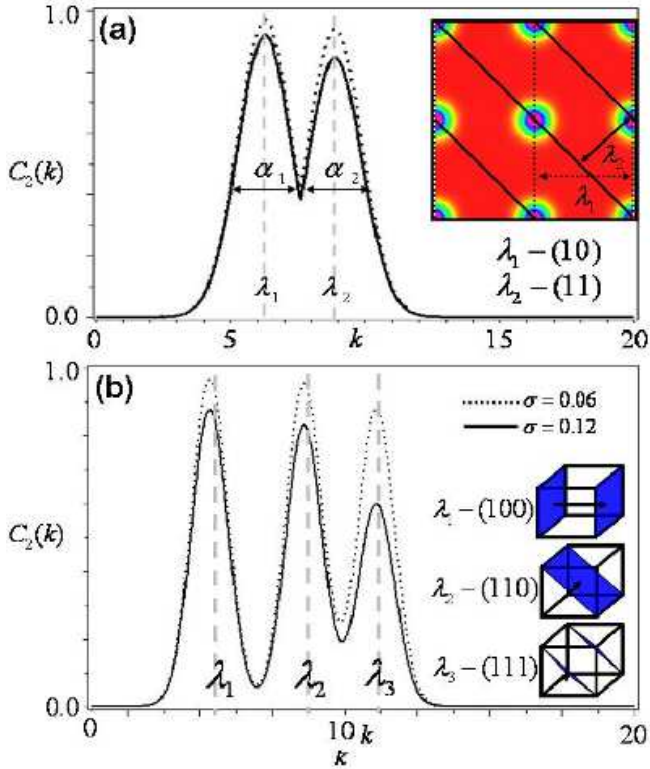}}
    \caption[]{\linespread{0.75} (a) Direct pair correlation function
      for a square lattice at two temperatures ($\sigma =0.82$, solid
      and $\sigma =0.5$, dotted, see text).  The two peaks in the
      correlation function each represent one of two planes in the
      unit cell with interplanar spacings $\lambda_1$ and $\lambda_2$
      (see inset).  (b) Correlation function for a simple cubic (SC)
      lattice at two different temperatures.  In 3D, a cubic lattice
      is represented by three planes in the unit cell.  The simple
      cubic cell has two planar spacings in common with a BCC lattice
      and requires the third peak to stabilize the structure.
    } \label{SQTRISCBCCCorr}
\end{figure}

Periodic structures emerge in the phase field crystal model through
the input of a direct correlation function into the excess free
energy. Here we shall construct correlation functions to create free
energy minima for arbitrary crystalline lattices by expanding the
free energy functional about a solid state reference density
\cite{Ebner1991}.  The original PFC model uses a single mode
correlation function; our correlation functions include multiple
modes, each of which corresponds to the spacing between atomic
planes within a particular unit cell. Figure \ref{SQTRISCBCCCorr}
shows two examples a the direct pair correlation function in
reciprocal space that generate stable square (a) and simple cubic
(SC) (b) lattices when the corresponding free energy is minimized.
The functions $C_2(k)$ are constructed in reciprocal space by
combining multiple peaks whose position, amplitudes and width are
determined as follows:

\begin{figure}[t]
    \centerline{\includegraphics*[width=3.4in]{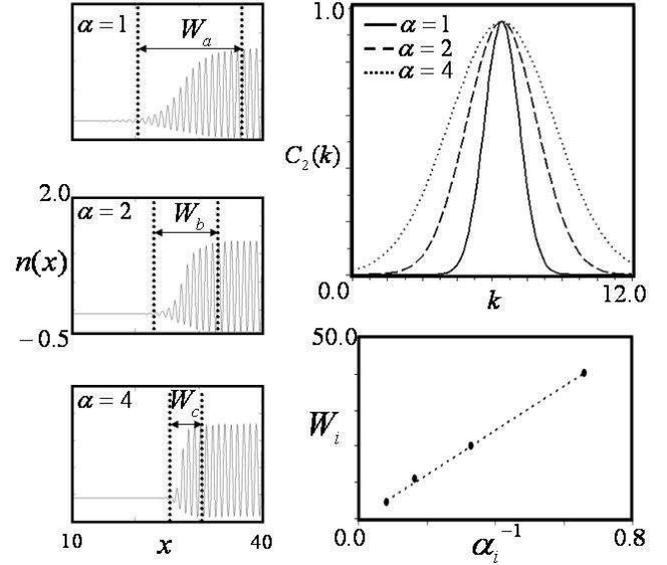}}
    \caption[]{\linespread{0.75} Effect of peak width $\alpha_i$ in
      the correlation functions on planar interface width $W_i$.
      Left: Three values of $\alpha_i$ and their resulting interface
      widths.  Top Right: Peak shapes corresponding the interface
      widths to the left.  Bottom Right: Dependence of simulated
      interface width on the fourier space correlation peak width (see
      text).  } \label{InterfaceWidth}
\end{figure}

The number of peaks in the correlation function and their reciprocal
space positions are determined by the unit cell of the desired
lattice.  Every plane within the unit cell produces a family of
peaks derived from the inter-planar spacings. For a 2D square
lattice, for instance, the unit cell contains two families of
planes, \{10\} and \{11\}, with spacings $\lambda_1$ and $\lambda_2$
as shown in the inset of Figure \ref{SQTRISCBCCCorr}(a). For a
perfect crystal, there are an infinite number of peaks located at
integer multiples of the wavevector $k_i =2\pi/\lambda_i$, where $i$
denotes the plane. For square and cubic lattices, it is sufficient
to keep only the lowest frequency mode for each family of these
peaks in the correlation function for each plane in the unit cell.

Temperature dependence enters the correlation function via
modulation of the peak heights through a Debye-Waller factor,
denoted $P_i$. As thermal motion causes atoms to vibrate about their
lattice positions, the reciprocal space lattice amplitudes fall off
by $e^{-\sigma^2k_i^2/2}$ where $\sigma$ is related to the
vibrational amplitude of the atoms and thus temperature. The peak
height is additionally influenced by the (dimensionless) atomic
density $\rho_i$ within a plane and the number of planes $\beta_i$
in each family. This effect can be seen when integrating out the
angular dependence in the reciprocal lattice. Here we modify the
Debye-Waller Factor to include both of these effects as $P_i =
\exp{[-\frac{\sigma^2k_i^2}{2\rho_i
      \beta_i}]}$. For example, in a square lattice the families of
$\{11\}$ and $\{10\}$ planes each consist of 4 sets of planes (ie.
the $\{11\}$ family contains $(11)$, $(\bar 11)$, $(1\bar 1)$, and
$(\bar 1 \bar 1)$) and therefore $\beta_{11} = \beta_{10} = 4$. The
$(11)$-plane has an atomic density $\rho_{11}=1/\sqrt{2}$ and the
$(10)$-plane has a density of $\rho_{10}=1$.

\begin{figure}[t]
    \centerline{\includegraphics*[width=3.4in]{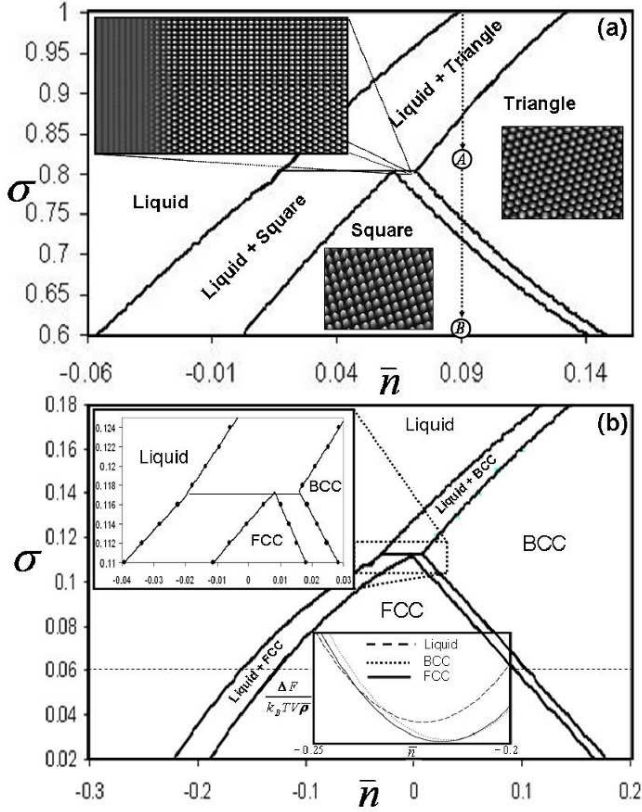}}
    \caption[]{\linespread{0.75}(a) Phase diagram for a square lattice
      correlation function showing coexistence between liquid, solid
      and triangular phases.  The inset contains stripes of square and
      triangular lattices in a liquid phase quenched to a temperature
      $\sigma = 0.79$. The square phase density was initialized to
      $n_{sq} = 0.067$ and the triangular phase density to $n_{tri} =
      0.076$.  The mean density of the system is set to $\bar{n} = 0.07$
      and the surface energy parameters are $\alpha_1 =\sqrt{2}$ and
      $\alpha_2=1$.  (b) Phase diagram for a FCC correlation
      function. The insets show the peritectic point and the energy
      curves for the liquid, FCC and BCC states at temperature of
      $\sigma = 0.06$.  } \label{PhaseDiagram}
\end{figure}

The peaks in the reciprocal space correlation function are
represented by Gaussians $\exp[-\frac{(k-k_i)^2}{2 \alpha_i^2}]$ of
finite width $\alpha_i$ rather than the $\delta-$peaks of a perfect
lattice. The parameter $\alpha_i$ provides a way to account for
changes in the free energy due to interfaces, defects and strain.
The effect of varying $\alpha_i$ on the width of a liquid-solid
interface is illustrated in Figure \ref{InterfaceWidth}. Increasing
the value of $\alpha_i$ leads to a decrease in the interface width,
which directly affects the surface energy as shown previously by
Majaniemi and Provatas \cite{Majaniemi2009}. In fact, the
relationship between the interface width and the peak width in the
correlation function is $1/\alpha_i \propto W_i$ as illustrated in
Figure~\ref{InterfaceWidth}.  This result can be arrived at by the
inverse fourier transform of $C_2(k)$ and agrees well with the
traditional phase field models, which incorporate surface energy
through square gradients in the order parameter. Note also that
$\alpha_i$ incorporates temperature dependence in the surface energy
through the Debye-Waller shift of the peak height.

In summary, each {\it family} of planes $i$ in the unit cell
contributes a peak to the direct pair correlation function of the
form \ben C_2(k)_i = e^{-\frac{\sigma^2k_i^2}{2\rho_i \beta_i}}
e^{-\frac{(k-k_i)^2}{2\alpha_i^2}}. \label{CorrPeak}\een The
correlation functions seen in Figure \ref{SQTRISCBCCCorr} are the
envelope of the superposition of all relevant peaks for the crystal
structure.  This construction described above provides a a general
and robust method by which to generate desired crystal structures,
which incorporate temperature dependence through the Debye-Waller
factor. This technique also allows for simple control over the
surface energy, which can be independently tuned to incorporate
anisotropy between planar facets.

By substituting the correlation functiond into the free energy
Eq.~(\ref{ModelEnergy}), the phase diagram is obtained at different
temperatures, using the free energy curves obtained for each phase.
The energy curves are calculated by an iterative relaxation
technique for each structure using the minimal kernel for the
structure of interest. For the liquid state, the energy is
calculated by imposing a constant density field, and the energy per
unit volume is calculated by numerically integrating
Eq.~(\ref{ModelEnergy}). For the solid state, the structures
corresponding to the correlation function are fit using a Gaussian
density field.  For example, for the correlation function of
Fig.~\ref{SQTRISCBCCCorr}(a), density fields for square and
triangular lattices with lattice spacings of 1 and $1/\sqrt{3}$ are
constructed and relaxed using Eq.~(\ref{Dyn:Dynamics}) to allow the
density peaks to obtain an amplitude corresponding to the local
minimum energy state. This process is repeated for a series of mean
densities to produce energy-density curves for each phase at a given
quench. An example of these curves is shown as an inset to
Fig.~\ref{PhaseDiagram}(b) for an FCC free energy functional. The
double tangent construction is performed for many quenches giving
the phase diagrams in Fig.~\ref{PhaseDiagram}(a) for a square
correlation function and in Fig.~\ref{PhaseDiagram}(b) for an FCC
correlation function.

\begin{figure}[t]
    \centering
    \begin{tabular}{ccc}
       \includegraphics[width=0.175\textheight]{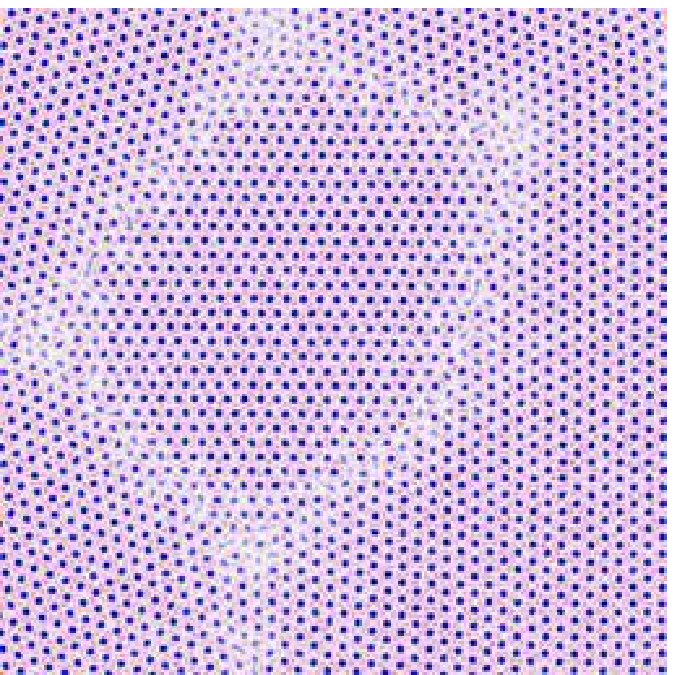} &
       \includegraphics[width=0.175\textheight]{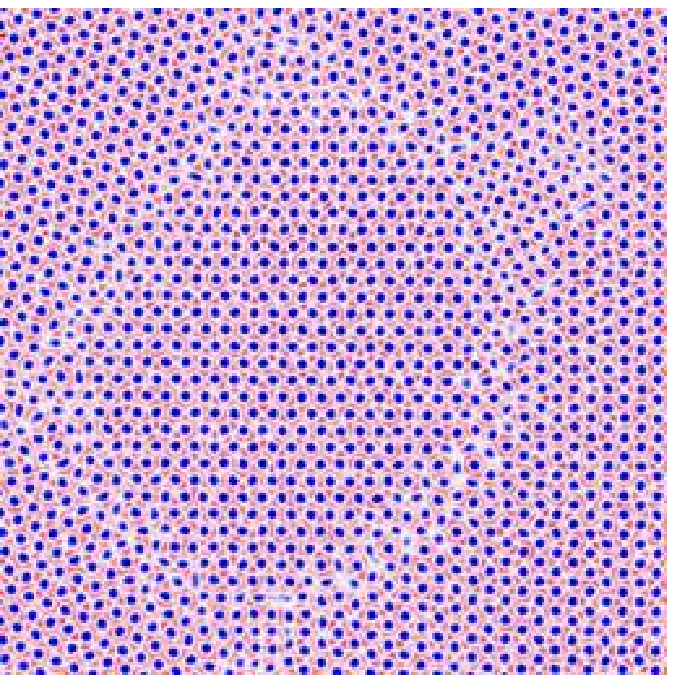} \\

       (a) time = 3
        &
        (b) time = 3.2
       \\
       \includegraphics[width=0.175\textheight]{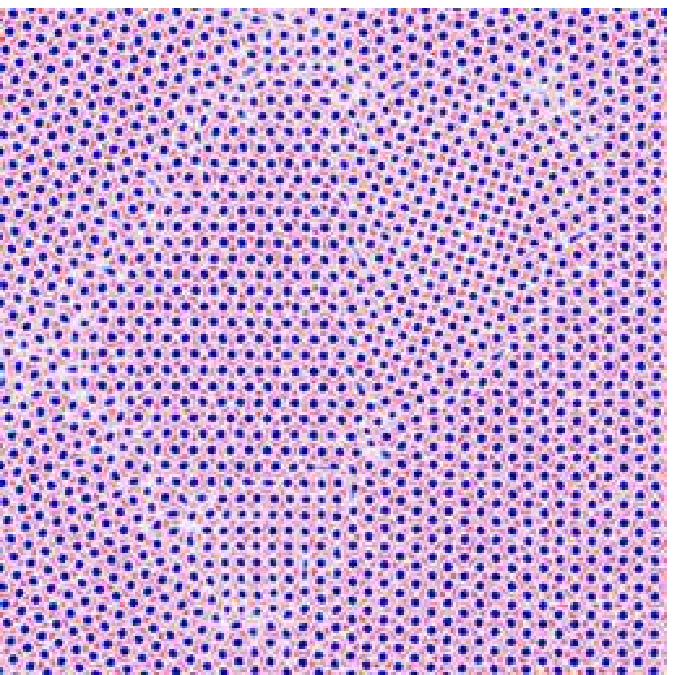} &
       \includegraphics[width=0.175\textheight]{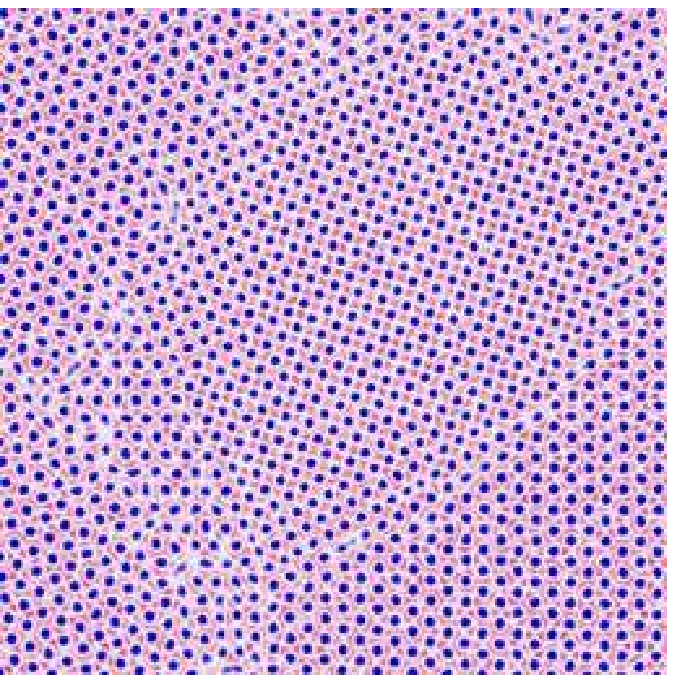}\\
       (c) time = 3.5   &  (d) time = 4
   \end{tabular}
    \caption[]{(Colour Online) 2D nucleation: from triangles to
      squares.  (a) A liquid with density $\bar{n} = 0.09$ is quenched
      into the triangular region of the phase diagram ($\sigma =
      0.82$), marked (A) in Fig.~\ref{PhaseDiagram}(a), produces
      grains with triangular symmetry.  (b-d) A second quench into the
      square portion of the phase diagram ($\sigma = 0.6$),
      marked (B) in Fig.~\ref{PhaseDiagram}(a), illustrating the
      nucleation and subsequent growth of square grains at the
      vertices of the triangular grains.} \label{structTransform}
\end{figure}

In both 2D and 3D, we find that the phase that possesses the
symmetry of the underlying correlation function emerges at low
temperatures and intermediate densities. As the temperature is
increased, the lowest frequency mode becomes dominant and for high
enough densities, the square (FCC) crystal transforms into a
triangular (BCC) crystal. For lower densities, the square (FCC)
crystal melts into a homogeneous liquid. Note that our model also
predicts a peritetic point in both dimensions, where two solid
phases coexist with a liquid.  This remarkable feature opens a
window into the study of structural phase transformations and
structural phase coexistence. Coexistence between FCC crystals with
the liquid phase is also easily obtained. Similarly, the correlation
function shown in Fig.~\ref{SQTRISCBCCCorr}(b) produces coexistence
of a liquid and SC phase.

Our new model is applied here to two important examples of
nonequilibrium solid state processese. The first is the growth of
two structurally different lamellae (stripes) into a liquid phase as
illustrated in the inset of Fig.~\ref{PhaseDiagram}(a). The square
phase is oriented such that the $\{10\}$ planes are in contact with
the liquid phase. Since each peak in the two particle correlation
function represents a single plane, the surface energy of the
interface is controlled by the width of the corresponding peak. The
square phase $\{10\}$-facet has its surface energy derived from the
width $\alpha_2$ of the second peak of the correlation function,
while the triangular phase facets are derived directly from the
width $\alpha_1$ of the first peak.  The square phase surface
energies can be tuned to isotropy by selecting the proper ratio of
the peak widths. Conversely, anisotropic surface energy can be
introduced by increasing or decreasing $\alpha_1$ while holding
$\alpha_2$ constant.  This effect also extends to the solid-solid
surfaces between boundaries of different structure (for example,
triangle-square boundaries) and misoriented solids of a structure
derived from multiple peaks (for example, the square structure). The
triangle structure is derived directly from a single peak and is
therefore isotropic for a given misorientation.  Small changes in
the strength of the anisotropy lead to small changes in the angle of
the solid liquid interfaces at the vertex point between the three
phases.

The second application of interest is the nucleation of a new
structural phase by a quench into the second phase region of the
phase diagram.  For illustrative purposes, the simple 2D case of
square-triangle structural transformations is used here. A liquid
with a homogenized density field can be quenched into the triangular
region of the phase diagram (see point A on the transformation path
in Figure ~\ref{PhaseDiagram}(a)) . Conserved noise in the dynamics
is added, and seeds of the triangular phase nucleate and are allowed
to coarsen for some time.  One such small grain is illustrated in
Fig.~\ref{structTransform}(a). After some time, the material can be
further quenched into the square portion of the phase diagram (see
point B).  Nucleation of the square phase and subsequent growth and
coarsening of the new phase is observed. Note that the nucleation
events happen at the vertex positions in the grain structures, as
observed in experiments \cite{Martin1997}.  Two such nucleation
points are shown in Fig.~\ref{structTransform}(b).  The orientation
of a nucleus and coherency strains can sometimes inhibit the growth
of the new phase. This can cause the new phase to be dominated by
another nucleation that is in a more favourable position to grow, an
effect illustrated in panels (c) and (d) of
Fig.~\ref{structTransform}.

We have introduced new CDFT/PFC free energy functionals for
efficient numerical study of solid state transformations.  In
contrast to earlier work \cite{Tupper2007}, we find that two-point
correlations are sufficient to generate stable cubic lattices, and
higher order terms in the expansion of $\Delta F_{ex}$ can still be
neglected. The correlation functions are systematically built up
from fundamental principles and desired crystallographic properties
of phases of interest at a finite temperature.  Our model captures a
peritectic transition as well as the nucleation and growth of
second-phase precipitates with different crystalline structures.

This work has been supported by the Natural Science and Engineering
Research Council of Canada (NSERC).


%

\end{document}